\documentstyle[12pt,axodraw]{article}
\begin{document}
\baselineskip 0.6cm

\begin{titlepage}

\begin{flushright}
UCB-PTH-01/41 \\
LBNL-49084 \\
\end{flushright}

\vskip 1.0cm

\begin{center}
{\Large \bf Gauge Coupling Unification from \\
  Unified Theories in Higher Dimensions}

\vskip 1.0cm

{\large
Lawrence J.~Hall and Yasunori Nomura
}

\vskip 0.5cm

{\it Department of Physics, University of California,
                Berkeley, CA 94720, USA}\\
{\it Theoretical Physics Group, Lawrence Berkeley National Laboratory,
                Berkeley, CA 94720, USA}

\vskip 1.0cm

\abstract{
Higher dimensional grand unified theories, with gauge symmetry breaking 
by orbifold compactification, possess $SU(5)$ breaking at fixed points, 
and do not automatically lead to tree-level gauge coupling unification. 
A new framework is introduced that guarantees precise unification --- 
even the leading loop threshold corrections are predicted, although 
they are model dependent. Precise agreement with the experimental result,
$\alpha_s^{\rm exp} = 0.117 \pm 0.002$, occurs only for a unique theory, 
and gives $\alpha_s^{\rm KK} = 0.118 \pm 0.004 \pm 0.003$.  Remarkably, 
this unique theory is also the simplest, with $SU(5)$ gauge interactions 
and two Higgs hypermultiplets propagating in a single extra dimension. 
This result is more successful and precise than that obtained from 
conventional supersymmetric grand unification, 
$\alpha_s^{\rm SGUT} = 0.130 \pm 0.004 \pm \Delta_{\rm SGUT}$.
There is a simultaneous solution to the three outstanding problems
of 4D supersymmetric grand unified theories: a large mass splitting 
between Higgs doublets and their color triplet partners is forced, proton 
decay via dimension five operators is automatically forbidden, and the 
absence of fermion mass relations amongst light quarks and leptons is 
guaranteed, while preserving the successful $m_b/m_\tau$ relation.
The theory necessarily has a strongly coupled top quark located on a 
fixed point and part of the lightest generation propagating in the bulk. 
The string and compactification scales are determined to be 
around $10^{17}$ GeV and $10^{15}$ GeV, respectively.} 

\end{center}
\end{titlepage}

\section{Introduction}

Weak scale supersymmetry not only provides a framework for electroweak
symmetry breaking, but also leads to a successful unification of gauge
couplings at extremely high energies.  If this picture of a 
supersymmetric desert is correct, then the scale of 
gauge coupling unification certainly heralds the threshold for 
some new physics. The desert will end with the appearance of some more 
unified theory.  What is this new physics just above the 
supersymmetric desert? There are two conventional answers.

From the bottom up viewpoint, a supersymmetric grand unified 
theory is the simplest interpretation of gauge coupling 
unification \cite{Dimopoulos:1981zb,Dimopoulos:1981yj}.
It gives an elegant explanation for charge quantization and the pattern 
of quark and lepton gauge quantum numbers. Furthermore, it can lead to a
reduction of parameters in the flavor sector leading to quark and lepton 
mass relations, and is a near perfect home for the see-saw mechanism for 
generating small, non-zero Majorana neutrino masses.  However, 
considerable obstacles are encountered in constructing simple and 
realistic 4D grand unified theories, even when low energy supersymmetry 
is included.  Chief amongst these are the mass splitting of Higgs 
doublets from their color triplet partners, proton decay induced by 
dimension five operators and the observed breaking of $SU(5)$ symmetry 
in the light quark and lepton masses. Indeed, the simplest supersymmetric 
$SU(5)$ theory is excluded by the limit on the proton lifetime.  
Furthermore, as we discuss in detail shortly, the prediction 
from gauge coupling unification is not in precise agreement with data.

The second conventional answer is that string theory is just above the
gauge coupling unification scale, without any energy interval with a
4D grand unified gauge symmetry. This top down approach is the only
serious contender for a quantum theory of gravity. It has brought
several new ideas relevant for the problems of gauge symmetry
breaking, doublet-triplet splitting and fermion mass relations 
\cite{Candelas:1985en,Dixon:jw}.  However, at present there is a barrier 
preventing a connection to the low energy domain: the problem of finding 
a consistent solution of string theory having a fully realistic 
low energy spectrum.

In a previous paper \cite{Hall:2001pg}, we have introduced a third 
possibility for the new unified physics which lies above the desert
and leads to gauge coupling unification: a higher dimensional 
grand unified theory compactified on an orbifold, with boundary condition 
breaking of the gauge symmetry.  This really is a new alternative 
--- there is no limit of the theory in which it contains the usual 
4D grand unified theories. It is not at all obvious that this new 
bottom up approach leads to gauge coupling unification, even at tree 
level, because there is local breaking of $SU(5)$ gauge invariance at the 
orbifold fixed points.  To recover gauge coupling unification, we find 
that the string scale must be considerably above the compactification 
scale. The large energy interval where physics is described by a higher 
dimensional grand unified theory distinguishes this scheme from the 
conventional string theory picture. It is remarkable that the three 
problems of 4D grand unified theories are all elegantly solved 
in this scheme. The orbifold projection of the unwanted color triplet 
Higgs zero mode, previously used in the context of string theory, 
transfers elegantly to the case of higher dimensional field 
theory \cite{Kawamura:2001ev}.  Lighter generations do not have unified 
fermion mass relations if they reside in the bulk, and dimension five 
proton decay from colored Higgsino exchange is automatically
removed by the form of the Higgsino mass matrix determined by higher
dimensional spacetime symmetry \cite{Hall:2001pg}.  We call this third 
framework for physics beyond the desert Kaluza-Klein (KK) 
grand unification.

In this paper we attempt to identify the leading candidate theory
within KK grand unification. We seek the simplest theory that gives 
a precise prediction for gauge coupling unification, without any 
significant dependence on unknown threshold corrections from 
high energies, and also solves the three problems of 4D grand 
unification.  This is clearly very ambitious.
In the case of 4D grand unification, unknown high energy threshold
corrections to gauge coupling unification are present, 
and are assumed to eliminate the discrepancy between theory and 
experiment. These corrections can arise from any number of complications 
to the spectrum of the theory.  In higher dimensions we find 
a contrasting result: a completely predictive and reliable
framework for gauge coupling unification follows from imposing  
a crucial new assumption of strongly coupled gauge interactions  
at the string scale. Remarkably, precise agreement with data implies 
an essentially unique theory, which is also the simplest. This again 
contrasts with the 4D case, where the simplest theory is excluded 
from proton decay.  The theory also provides a new arena for an 
understanding of quark and lepton masses: at least some of the hierarchy 
amongst fermion masses arises from the large volume of the bulk. 
Furthermore, the theory gives a clearly successful correlation: heavier 
fermions should display $SU(5)$ mass relations while lighter ones 
should not.

The key ingredient to uncovering this higher dimensional unified theory 
is gauge coupling unification, so we start by reviewing the situation 
in 4D grand unified theories \cite{Georgi:1974sy,Georgi:1974yf}. 
The electroweak gauge couplings are now so well measured, that we choose 
to input them from data and give a prediction for the QCD coupling 
at the scale of the $Z$ mass, which has a measured value 
$\alpha_s^{\rm exp} = 0.117 \pm 0.002$ \cite{Groom:2000in}.  
Assuming the standard model holds to extremely high energies of 
order $10^{15}$ GeV, the grand unified prediction is
\begin{equation}
  \alpha_s^{\rm GUT} = 0.077 \pm \Delta_{\rm GUT}.
\label{eq:alphas.sm}
\end{equation}
The physics at the weak scale and in the desert up to the 
unified mass scale is assumed known, so that this part of the calculation 
has essentially no uncertainly. The quantity $\Delta_{\rm GUT}$ arises 
from the physics at the unified mass scale, which is not well known. 
This threshold correction must be very large, correcting the leading 
order prediction by 50\%. One therefore concludes that in this theory
there is no precise prediction of the QCD coupling: the threshold
correction depends on some continuous parameter of the unified theory
which simply has to be chosen to give the experimental result.

The situation is greatly improved in supersymmetric grand unified 
theories \cite{Dimopoulos:1981zb,Dimopoulos:1981yj}. Superpartners 
at the TeV scale modify the radiative corrections to the gauge 
couplings, leading to the prediction \cite{Langacker:1995fk}
\begin{equation}
  \alpha_s^{\rm SGUT} = 0.130 \pm 0.004 \pm \Delta_{\rm SGUT}.
\label{eq:alphas.susy}
\end{equation}
The first uncertainty arises from variations in the superpartner
spectrum at the TeV scale, while the second uncertainty arises from
the unknown spectrum of states at the unification scale.\footnote{
At first sight it appears that this prediction is accurate at the 10\% 
level.  However, this understates the significance of the result. 
Viewing the prediction as a correlation in the plane of the QCD coupling 
and weak mixing angle, one finds the accuracy to be at about the 1\% level
\cite{Hall:1996gq}. This is then the most significant prediction of 
any of the 18 free parameters of the standard model, explaining 
the wide attention it has received.} 
While a non-zero threshold correction is necessary for agreement with data, 
one can take the viewpoint that such corrections are typically expected.  
This prediction is a crucial part of the motivation for the 4D grand 
unification paradigm, but a purist might still object that in any particular
model the correction $\Delta_{\rm SGUT}$ depends on unknown continuous
parameters, so that formally there is no prediction. In the minimal
$SU(5)$ supersymmetric unified theory there is only one such parameter,
which can be taken to be the mass of the heavy color triplet partners
of the Higgs doublets. Unfortunately, the sign of the correction is
such that the value of the Higgs triplet mass needed for the
prediction for the QCD coupling to agree with data is less than the
unified mass scale, and is excluded by the experimental limit on the
proton lifetime. There are no known models where $\Delta_{\rm SGUT}$ 
can either be successfully predicted from theory or constrained from
independent data. As the experimental data on gauge couplings has 
improved, the requirement for a large value of $\Delta_{\rm SGUT}$
has become more pronounced. For example, if superheavy 
${\bf 5} + \bar{\bf 5}$ chiral multiplets of $SU(5)$ are added, a unit 
logarithmic mass splitting between doublet and triplet components gives
$\Delta_{\rm SGUT} \simeq 0.003$. The data requires several such 
multiplets, a large Casimir, or a large mass splitting.

In higher dimensional gauge theories, the grand unified symmetry can be 
broken by orbifold boundary conditions. At first sight this is a
disaster. The orbifold contains fixed sub-spaces on which the unified
symmetry is explicitly broken. Local gauge kinetic terms on these
sub-spaces lead to a tree-level violation of gauge coupling unification, 
so that the prediction for the QCD coupling is completely lost. 
In Ref.~\cite{Hall:2001pg} we overcame this difficulty by requiring the
bulk to have a large volume, and introduced a new framework for gauge 
coupling unification in higher dimensional unified 
theories, which we pursue further here.  The improvements from 
supersymmetry are retained, with superpartner contributions to the 
evolution of gauge couplings from the weak scale to some high energy 
scale $M_c$. At this scale the picture is greatly altered by the opening 
up of extra spatial dimensions: one expects threshold corrections at 
$M_c$ and power-law running of the three gauge couplings above $M_c$ 
up to the string scale $M_s$.  The crucial new ingredient follows from 
the restricted set of unified gauge transformations in the bulk, which 
are determined by the orbifold boundary conditions \cite{Hall:2001pg}. 
This bulk symmetry ensures that the leading, power-law evolution is 
$SU(5)$ symmetric. However, fixed points in the bulk do not respect 
the full unified gauge symmetry and lead to an additional evolution: 
a non-universal logarithmic running of the standard model gauge 
couplings in the energy region above $M_c$.  

The prediction for the QCD coupling therefore involves three terms: 
supersymmetric evolution from $M_Z$ to $M_c$ involving a very large 
logarithm $\ln(M_c/M_Z)$, threshold contributions at $M_c$ and $M_s$, 
and a moderately large logarithmic term proportional to $\ln(M_s/M_c)$ 
originating from the KK towers. It is this structure that allows 
a completely predictive framework for gauge coupling unification.  
Requiring the gauge coupling to be strongly coupled at the string scale 
suppresses unknown contribution from ultraviolet physics to a negligible 
level \cite{Nomura:2001tn}.  The discrepancy between the usual 
supersymmetric prediction, Eq.~(\ref{eq:alphas.susy}), and data is then 
provided by the second KK logarithm, which is smaller than the first 
supersymmetric logarithm, but larger than the non-logarithmic threshold 
corrections.  It is remarkable that even though there are many theories 
of this type, depending on the number of compact dimensions, the nature 
of the orbifold and the gauge group, we find that only one gives precise 
agreement with data.  This theory is also the minimal possibility, with 
gauge group $SU(5)$ in 5D broken to the standard model on the orbifold 
$S^1/Z_2$, giving
\begin{equation}
  \alpha_s^{\rm KK} = 0.118 \pm 0.004 \pm 0.003,
\label{eq:alphas.kk}
\end{equation}
where the first error bar is the uncertainty from the superpartner 
spectrum, which can be eliminated by future measurements, and the 
second error is from residual uncertainties from physics at $M_s$. 
It is important that the theory does not contain any free parameter 
that can be used to adjust the prediction for the QCD coupling.
The masses and couplings of the unified scale particles are all fixed 
by the orbifold boundary conditions, and the ratio $M_s/M_c$ is 
determined by the strong coupling requirement.  The essential features 
of this theory are illustrated in Fig.~\ref{fig:frame}.

In section \ref{sec:framework} we define a framework which leads to 
a precise prediction for the QCD coupling, and find that the 
compactification which breaks the unified gauge symmetry involves 
at most two extra dimensions. In section \ref{sec:orbifold} we show 
that this framework precisely accounts for the data only if this 
compactification is on $S^1/Z_2$ and the unified gauge group is $SU(5)$. 
The size of the theoretical uncertainties are discussed. 
In section \ref{sec:su5} a fully realistic $SU(5)$ model is explored,
concentrating on the solution to the three outstanding problems of 4D
supersymmetric grand unified theories: the splitting between doublet
and triplet Higgs masses, dimension five proton decay and the absence of
$SU(5)$ mass relations for the first two generations. We investigate
to what extent fermion masses and mixings can be understood from
locality in the bulk, and briefly mention possible signatures from
dimension six proton decay, which occurs only via flavor mixing. 
In section \ref{sec:string} we comment on the possible origin of 
our theory from string theory.  Conclusions are drawn in 
section \ref{sec:concl}.

\section{The Framework}
\label{sec:framework}

KK grand unification provides a third possibility for physics above
the supersymmetric desert. Here we push this idea further and 
propose a completely predictive framework for gauge coupling 
unification, which follows from the additional assumption that the 
theory is strongly coupled at the string scale. This assumption, 
which we find quite plausible, ensures that threshold corrections from 
the string scale are highly suppressed, and determines the size of the
leading loop corrections below the string scale.
Our framework is defined by the following five elements:
\begin{itemize}
\item 
 We introduce two mass scales $M_c$ and $M_s$, rather than a single 
 unification mass scale.
\item
 The scale, $M_c = 1/R$, characterizes the size of $d$ extra spatial
 dimensions. The structure of this $d$ dimensional compact space is
 chosen so that the framework leads to a precise prediction for gauge
 coupling unification without significant sensitivity to unknown 
 ultraviolet physics.
 (We will find that $M_c$ must be taken very large, of order $10^{15}$ GeV, 
  but it differs from the usual unification mass scale.)
\item
 The effective theory above $M_c$ is a higher dimensional
 grand unified theory with gauge group $G$. This gauge group is broken
 at $M_c$, by boundary conditions of the extra spatial
 dimensions, to the standard model gauge group (together with
 possible extra factors).
\item
 The effective theory below $M_c$ is the minimal supersymmetric
 standard model (MSSM).
\item
 $M_s$ is the scale at which the effective higher dimensional theory 
 is embedded into a more fundamental theory such as string theory. 
 We identify $M_s$ as the scale where gauge interactions of $G$ become 
 strongly coupled: $C(g^2/16 \pi^2)(M_s/M_c)^d \simeq 1$, where $C$ is 
 a group theoretical factor appearing in the loop expansion.
 (For example, in the case of $G = SU(N)$, $C \simeq N$.)
 Since $g$ evolves slowly up to energies very close to $M_s$,
 we may estimate $M_s$ by taking $g$ to be the 4D gauge coupling at 
 the scale $M_c$, $g \simeq 0.7$, giving $(M_s/M_c)^d \simeq 300/C$. 
 (Note that $d$ cannot be taken too large, otherwise the two masses 
 $M_c$ and $M_s$ do not represent different scales.)
\end{itemize}

This framework appears to be very broad, encompassing many
possibilities for $G$, $d$ and the compact space. From the viewpoint
of gauge coupling unification it is convenient to divide the set of
higher dimensional unified theories into four types, I --- IV. 
In Fig.~\ref{fig:types} the evolution of the three standard model gauge 
couplings $g_i$ are illustrated for each type of theory, showing the 
behavior in the energy intervals $M_Z$ to $M_c$ and $M_c$ to $M_s$.
In general, the prediction for gauge coupling unification depends not
only on the zero modes, but on the entire towers of KK modes. We
therefore discuss the gauge symmetry structure of the entire higher
dimensional theory. 
\begin{figure}
\begin{center}
\begin{picture}(450,350)(-130,-20)
%
%
  \Text(-30,190)[t]{a) Type I theories}
  \Text(60,318)[b]{$\ln\mu$}
  \Text(-105,330)[rb]{$\eta_i$}
  \Text(-110,307)[rt]{$\eta_1$}
  \Line(-30,310)(-120,265)
  \Text(-110,264)[rt]{$\eta_2$}
  \Line(-30,310)(-120,233)
  \Text(-110,232)[rt]{$\eta_3$}
  \LongArrow(-100,190)(-100,330)
  \LongArrow(-120,310)(60,310)
  \Line(-30,306)(-30,314) \Text(-30,320)[b]{$M_c$}
  \Line(20,306)(20,314) \Text(20,320)[b]{$M_s$}
  \DashLine(-30,310)(-30,250){3}
%
%
  \Text(220,190)[t]{b) Type II theories}
  \Text(310,318)[b]{$\ln\mu$}
  \Text(145,330)[rb]{$\eta_i$}
  \Text(140,307)[rt]{$\eta_1$}
  \CArc(300,216)(99,108,144)
  \Line(220,274)(130,229)
  \DashLine(220,274)(300,314){2}
  \Text(140,239)[rb]{$\eta_2$}
  \CArc(300,112)(200,99,114)
  \Line(220,295)(130,218)
  \DashLine(220,295)(255,330){2}
  \Text(140,217)[rt]{$\eta_3$}
  \LongArrow(150,190)(150,330)
  \LongArrow(130,310)(310,310)
  \Line(220,306)(220,314) \Text(220,320)[b]{$M_c$}
  \Line(270,306)(270,314) \Text(270,320)[b]{$M_s$}
  \DashLine(220,310)(220,250){3}
%
%
  \Text(-30,0)[t]{c) Type III theories}
  \Text(60,128)[b]{$\ln\mu$}
  \Text(-105,140)[rb]{$\eta_i$}
  \Text(-110,117)[rt]{$\eta_1$}
  \Line(20,120)(-120,50)
  \Text(-110,60)[rb]{$\eta_2$}
  \Line(20,120)(-120,0)
  \Text(-110,13)[rb]{$\eta_3$}
  \LongArrow(-100,0)(-100,140)
  \LongArrow(-120,120)(60,120)
  \Line(-30,116)(-30,124) \Text(-30,130)[b]{$M_c$}
  \Line(20,116)(20,124) \Text(20,130)[b]{$M_s$}
  \DashLine(-30,120)(-30,60){3}
%
%
  \Text(220,0)[t]{d) Type IV theories}
  \Text(310,128)[b]{$\ln\mu$}
  \Text(145,140)[rb]{$\eta_i$}
  \Text(140,117)[rt]{$\eta_1$}
  \Line(270,120)(220,102)
  \Line(220,102)(130,57)
  \DashLine(220,102)(270,127){2}
  \Text(140,67)[rb]{$\eta_2$}
  \Line(270,120)(220,93)
  \Line(220,93)(130,16)
  \DashLine(220,93)(260,127){2}
  \Text(140,29)[rb]{$\eta_3$}
  \LongArrow(150,0)(150,140)
  \LongArrow(130,120)(310,120)
  \Line(220,116)(220,124) \Text(220,130)[b]{$M_c$}
  \Line(270,116)(270,124) \Text(270,130)[b]{$M_s$}
  \DashLine(220,120)(220,60){3}
\end{picture}
\caption{The running of the difference of the three gauge couplings, 
 $\eta_i \equiv \alpha_i^{-1} - \alpha_1^{-1}$, below and above $M_c$.
 In type I and III theories, the gauge couplings unify at $M_c$ and 
 $M_s$, respectively.  In type II theories, the successful prediction 
 is typically destroyed by the non-universal tree-level and power-law 
 running contributions to $\eta_i$. (In the case that the tree-level 
 contributions are small, the situation is similar to that 
 in type I theories.)
 In type IV theories, a naive extrapolation of the low energy 
 gauge couplings leads to an approximate unification at a scale between 
 $M_c$ and $M_s$, giving a small deviation from the case of 
 single-scale unification.}
\label{fig:types}
\end{center}
\end{figure}

Compactification is obtained by imposing a set of identifications 
on the space of the extra dimensions: $y \rightarrow k(y)$. 
To break the unified gauge symmetry, while leaving an unbroken
subgroup, the gauge fields are chosen to be even or odd under each
such identification: $A_\mu^{a_\pm}(x,y) \rightarrow \pm
A_\mu^{a_\pm}(x,k(y))$. This set of gauge fields results only if the
underlying gauge symmetry of the theory has
the form: $\xi^{a_\pm}(x,y) \rightarrow \pm \xi^{a_\pm}(x,k(y))$, 
which we refer to as restricted gauge symmetry \cite{Hall:2001pg}. 
At a typical location in the bulk, all gauge parameters $\xi^a$ are
non-zero, so that the bulk is $G$ invariant.

In type I theories the $d$ dimensional space is a manifold. This 
means that all identifications are freely acting:
there is no point in the bulk for which $k(y) = y$. 
The full gauge invariance of $G$ acts locally at every point in the 
manifold, so that the gauge couplings unify at $M_c$, as shown in 
Fig.~\ref{fig:types}a. There are no $G$-violating effects at distances 
below $R$.  The prediction for the QCD gauge coupling is the usual one 
for supersymmetric unification, Eq.~(\ref{eq:alphas.susy}). However, 
the threshold corrections $\Delta_{\rm SGUT}$ from $M_c$ can now be 
computed. They arise from the KK excitations of the Higgs and gauge 
multiplets and are much too small to explain the difference between 
the prediction, $0.130 \pm 0.004$, and the data, $0.117 \pm 0.002$. 
If extra bulk multiplets are added, the boundary conditions on the 
manifold will lead to additional zero modes which are not in complete 
$SU(5)$ multiplets, with disastrous results for gauge coupling 
unification.  Theories compactified on a manifold could become 
interesting if experiments find the spectrum of superpartners to be 
far from that of theoretical expectations, so that weak scale threshold 
corrections bring the prediction for standard supersymmetric unification 
into agreement with data. Even in this case, however, obtaining 
a realistic low energy theory with chiral fermions may be difficult 
on such smooth spaces.

The remaining theories are all compactified on orbifolds, and easily 
lead to low energy chiral theories. Orbifolds result when there is at 
least one identification having $k(\bar{y})=\bar{y}$, and we call 
$\bar{y}$ a fixed sub-space or a brane. The gauge parameters and 
associated gauge fields which are odd under this identification vanish 
on the fixed sub-space, $\xi^{a_-}(\bar{y}) = 0$. On $\bar{y}$, the 
gauge symmetry is broken from $G$ to a subgroup $H$ generated by 
$\xi^{a_+}$. Hence, the restricted gauge symmetry, resulting from the 
orbifold boundary conditions, allows local $G$ violation 
\cite{Hall:2001pg}.  Since we are using an effective field theory 
viewpoint, the most general set of $H$ invariant operators occurs 
on the orbifold fixed sub-spaces, leading to explicit, local breaking
of $G$. In particular, there are fixed sub-spaces giving kinetic energy 
operators for the standard model gauge fields with non-unified 
coefficients $1/\tilde{g}_i^2$.  At first sight, such ``non-universal 
fixed sub-spaces'' ruin gauge coupling unification.  However, we find 
that this need not be the case, although these branes do play 
an important role for gauge coupling unification, at both tree and 
quantum levels.

To see the effect of local $G$ violation, we first consider the 
effective action at the scale $M_s$.  No matter what physics occurs above 
$M_s$, the restricted gauge symmetry ensures that the ($4+d$)-dimensional 
bulk is $G$ symmetric and all $G$-violating effects appear only on 
$G$-violating branes. Therefore, the most general form for the gauge 
kinetic energy is given by
\begin{equation}
  S = \int d^4x \; d^dy \; 
    \biggl[ \frac{1}{g_{4+d}^2} F^2 + 
    \delta^{(d-\delta)} (y - \bar{y}) \frac{1}{\tilde{g}_i^2} F_i^2 \biggr],
\label{eq:gaugekinops}
\end{equation}
where the first term is a $G$-invariant bulk gauge kinetic energy, 
while the second term represents non-unified kinetic operators 
located on a non-universal brane of dimensions $4+\delta$.
(In general, there are contributions from several non-universal branes 
and also from $G$-symmetric branes.)  The standard model gauge couplings 
in the equivalent 4D KK theory are obtained by integrating 
over the $d$ extra dimensions:
\begin{equation}
  \frac{1}{g_i^2} = \frac{V}{g_{4+d}^2} + \frac{V'}{\tilde{g}_i^2},
\end{equation}
where $V$ is the volume of the bulk, and $V'$ the volume of the
non-universal brane ($V'=1$ if $\delta=0$).  Now, since the theory 
is assumed to be strongly coupled at $M_s$, both bulk and brane 
gauge couplings are reliably estimated as $g_{4+d} \approx \tilde{g}_i 
\approx 4\pi$ in units of $M_s$.\footnote{
They are estimated, for example, by considering loop diagrams 
in the equivalent 4D KK theory.  In the 4D picture, the bulk term gives 
gauge kinetic terms with KK momentum conservation, while the brane ones 
give terms with KK momentum violation.  After diagonalizing these kinetic 
terms, the gauge couplings among KK towers are obtained.
Requiring that contributions from all loop diagrams become comparable 
at the scale $M_s$ (i.e. the theory is strongly coupled at $M_s$), we 
obtain the result $g_{4+d} \approx \tilde{g}_i \approx 4\pi$ in units of 
$M_s$, neglecting group theoretical factors of order unity.}
Thus the tree-level values of the 4D gauge couplings at $M_s$ are given by
\begin{equation}
  \frac{1}{g_i^2} \approx \frac{V}{16 \pi^2} + c_i\frac{V'}{16 \pi^2}, 
\end{equation}
where $c_i \approx 1$ represent non-universal coefficients.
Here and below, $V$ and $V'$ are given in units of $M_s$.
The requirement of our framework that gauge coupling unification 
is insensitive to unknown ultraviolet physics
translates to the simple requirement that $V'/V$ is sufficiently
small for all non-universal branes.\footnote{
In general, if the bulk and brane gauge kinetic terms are comparable 
at the string scale (i.e. $g_{4+d} \approx \tilde{g}_i$ in units of 
$M_s$), the non-universal contribution to the zero mode gauge couplings 
from the brane terms is suppressed compared with the universal bulk 
contribution by a factor of $V'/V$.  A formal understanding of this fact 
is given as follows.  Since the non-universal brane kinetic operators, 
$\delta^{(d-\delta)} (y - \bar{y}) F_i^2$, have higher mass dimensions 
(i.e. they are more irrelevant in the Wilsonian sense) than the 
$G$-preserving bulk kinetic term, $F^2$, by an amount $d-\delta$ 
corresponding to the dimension of the delta function, 
the effect of the former is suppressed relative to the latter 
at lower energies $\mu$.  The suppression factor is given by 
$(\mu/M_s)^{d-\delta}$, which exactly gives the transverse volume 
factor $V'/V$ at the compactification scale, 
$\mu \approx M_s V^{-1/d} \approx M_s V'^{-1/\delta}$, that is, 
the relevant scale for the zero mode gauge fields.}
This becomes easier to satisfy as the dimension of the branes is 
reduced.  Furthermore, since the value of the unified 4D gauge coupling 
is of order unity, the volume of the bulk is large in fundamental 
units: $V \approx 100$.

Having obtained gauge coupling unification at tree level at $M_s$, 
we turn to the quantum effects below $M_s$ that result 
from non-universal branes.  Consider first the 
case of such a brane with dimension $\delta >0$. At one loop, the scaling 
from $M_s$ to $M_c$ can give a non-universal correction to $1/g_i^2$
by an amount $(1/16 \pi^2)(M_s/M_c)^\delta \approx V' /16 \pi^2$.
Thus the requirement that the tree-level unification of gauge couplings 
is insensitive to unknown physics at $M_s$ also guarantees insensitivity 
at the loop level to non-universal branes of $\delta > 0$. The relative 
power-law running of $g_i$ induced by such branes is found, perhaps
surprisingly, to be unimportant. Type II theories are defined as those 
having non-universal branes with $\delta>0$, but none with $\delta = 0$. 
These theories typically do not satisfy the condition of ultraviolet 
insensitivity, and the successful prediction is destroyed by the 
non-universal tree-level and power-law running contributions.  
Even in the case that the condition is satisfied, they predict 
$\alpha_s = 0.130 \pm 0.004$ like type I theories on manifolds, 
and hence are excluded for conventional superpartner spectra.

The remaining theories are those which possess $\delta = 0$
non-universal fixed points. The crucial aspect of these fixed points
is that they induce a non-universal logarithmic running of the gauge
couplings; from $M_s$ to $M_c$, $1/g_i^2$ is corrected 
by an amount $(1/16 \pi^2) \ln(M_s/M_c)$.  Since the non-universal 
tree correction factor is $1/ 16 \pi^2$, the loop contribution dominates 
by $\ln(M_s/M_c)$. We find a remarkable result: only in the case of 
$\delta = 0$ are the relative corrections to the gauge couplings
dominated by loop rather than tree effects. Furthermore, since the
loop effects are logarithmic, they can be reliably computed in the
effective theory. The unknown contributions from $M_s$ are suppressed
relative to this calculable term by $1/\ln(M_s/M_c)$. 

We thus concentrate on theories having non-universal branes with 
$\delta = 0$, in the hope that the logarithmic running above $M_c$ will 
lead to a precise agreement of gauge coupling unification with data.
For the case $d>2$, this does not happen. To obtain a
supersymmetric theory below $M_c$, the unified theory in $4+d$
dimensions must be supersymmetric. Supersymmetry in high dimensions is
very constraining, corresponding to several supersymmetries in the 4D
picture. If $d>2$, each excited KK level of the equivalent 4D
theory has 4D supersymmetry with $N \geq 4$, and hence does not
contribute to the running of $g_i$. The evolution of $g_i$ above $M_c$
is only due to the zero modes, which by construction are those of the
MSSM with $N=1$, and hence is the same as the evolution below
$M_c$. For $d>2$ gauge coupling unification mimics the conventional
supersymmetric case, with $M_s$ as the unification scale. 
This also occurs for $d=1,2$ if the higher dimensional theory has more
than one supersymmetry. These theories we call type III, and the
coupling evolution is shown in Fig.~\ref{fig:types}c.

Therefore, we find that a precise and successful prediction for 
gauge coupling unification is possible only for a 5D or 6D unified 
theory, with minimal amount of supersymmetry, compactified on an 
orbifold having non-universal fixed points, but not non-universal 
fixed lines.  (Non-universal fixed lines in 6D give too large unknown 
corrections to $1/g_i^2$, of size $1/4\pi$.)  The picture of gauge 
unification for these type IV theories is shown in Fig.~\ref{fig:types}d.
This result is consistent with the requirement that $M_s$ and $M_c$ 
are well separated: in the minimal case of $G = SU(5)$, for instance, 
there are about $60$ ($8$) KK excitations of each zero mode 
for $d=1$ ($2$), so that there is a substantial energy
interval having physics described by a higher dimensional field
theory. This is quite unlike the case of string theory
compactified on a 6D compact space with comparable sizes for the
six dimensions. For strongly coupled string theory there are 2 KK 
excitations for each zero mode \cite{Horava:1996qa},
and fewer for the case of weak coupling \cite{Candelas:1985en,Dixon:jw}.

The leading correction to gauge coupling unification arises from
the logarithmic contribution to the running above $M_c$, as shown in
Fig.~\ref{fig:types}d, and is proportional to $\ln(M_s/M_c)$. 
There are further corrections which are not logarithmically enhanced: 
threshold corrections from $M_s$ and $M_c$.  Those from $M_s$, 
representing unknown physics from higher energies, correct $1/g_i^2$ 
by $1/16 \pi^2$, leading to an uncertainty in $\alpha_s$ of $\pm 0.002$.
As for threshold corrections from $M_c$, they arise from any multiplets 
that are $SU(5)$ split in any particular model.  In the higher 
dimensional picture, they are represented by non-local operators 
spread out in the extra dimensions.  While there are no contributions 
from matter, even if they arise from several multiplets of $G$, there 
are contributions from the multiplets of $G$ which contain the zero mode 
Higgs doublets and standard model gauge fields. These contributions are 
computed and included in our result, even though they correct $1/g_i^2$ 
only by $1/16 \pi^2$ and thus are small. There can be
no other split multiplets with zero modes, otherwise physics below
$M_c$ would not be described by the MSSM, and these additional zero 
modes would ruin the success of supersymmetric unification, 
Eq.~(\ref{eq:alphas.susy}). However, there could be extra bulk multiplets 
with brane mass terms giving the zero modes a mass of order $M_c$. 
We assume that the dominance of the $\ln(M_s/M_c)$ correction is not 
spoiled by a large number of such multiplets.

\section{Determining the Orbifold and Gauge Group}
\label{sec:orbifold}

In this section, we calculate the correction to $\alpha_s$ proportional 
to $\ln(M_s/M_c)$ for type IV theories and determine the structure 
of the theory above $M_c$ using the experimental value of $\alpha_s$.
We first derive a general formula for the KK tower contribution to 
$\alpha_s$.  In type IV theories, the low-energy values for the three 
gauge couplings are given by 
\begin{equation}
  \frac{1}{g_i^2}(\mu) = \frac{1}{g_*^2} 
    + \sum_{k=1}^{d} \frac{c_k b^{(k)}}{8 \pi^2 k} 
      \left[ \left(\frac{M_s}{M_c}\right)^{k} - 1 \right]
    + \frac{\tilde{b}_i}{8 \pi^2} \ln\frac{M_s}{M'_c}
    + \frac{b'_i}{8 \pi^2} \ln\frac{M'_c}{\mu} 
    + \frac{\Delta_i}{8 \pi^2},
\label{eq:running}
\end{equation}
where $b'_i$ are the $\beta$-function coefficients for the MSSM, 
$(b'_1, b'_2, b'_3) = (33/5, 1, -3)$, and $b^{(k)}$ and $\tilde{b}_i$ 
those for the theory above $M_c$; $g_*$ is the unified gauge coupling 
at $M_s$, $c_k = \pi^{k/2}/\Gamma(1+k/2)$, and $\Delta_i$ represent 
the effects of threshold corrections from $M_s$ and $M_c$. 
The power-law terms proportional to $(M_s/M_c)^k-1$ come from the running 
of the bulk gauge coupling (or gauge kinetic terms on $G$-symmetric 
fixed lines) and thus must be universal, while the term proportional 
to $\ln(M_s/M'_c)$ comes from the running of gauge kinetic terms localized 
on the (non-universal) fixed points and can depend on $i$ 
\cite{Hall:2001pg}.  Here, we have matched the logarithmic contribution 
in higher dimensions to that in 4D at the scale $M'_c = M_c/\pi$, 
which represents the length scale of extra dimensions.
This is the natural scale for the matching, as indicated by 
summing up leading-log contributions from KK towers \cite{Nomura:2001mf}.
Using the above equations, we obtain one relation among $g_i$'s 
\begin{equation}
  \frac{1}{g_3^2} = \frac{12}{7}\frac{1}{g_2^2} 
    - \frac{5}{7}\frac{1}{g_1^2}
    + \frac{\tilde{b}}{8\pi^2} \ln\frac{M_s}{M'_c} 
    + \frac{\Delta}{8\pi^2},
\label{eq:KK-relation}
\end{equation}
at any scale $\mu$ ($< M'_c$). 
Here, $\tilde{b}$ and $\Delta$ are defined by
\begin{equation}
  \tilde{b} = \tilde{b}_3 
    - \frac{12}{7}\tilde{b}_2 + \frac{5}{7}\tilde{b}_1,
\label{eq:tilde-b}
\end{equation}
and $\tilde{b},\tilde{b}_i \rightarrow \Delta,\Delta_i$.
Note that the dependence on $b^{(k)}$ drops out since the power-law 
pieces are universal.  

Suppose we compute $\alpha_s$ from the observed electroweak gauge 
couplings $g_1$ and $g_2$, using Eq.~(\ref{eq:KK-relation}).  
Then, the obtained value $\alpha_s^{\rm KK}$ is in general different from 
the value $\alpha_s^{\rm SGUT,0}$ obtained in the case where the 
couplings unify at a single scale $M_u$ without any threshold correction 
(which corresponds to setting $M_s = M_c = M'_c = M_u$ and 
$\Delta_i = 0$ in Eq.~(\ref{eq:running})).  The difference 
$\delta\alpha_s \equiv \alpha_s^{\rm KK} - \alpha_s^{\rm SGUT,0}$ 
is given by
\begin{equation}
  \delta\alpha_s = -\frac{1}{2\pi} \alpha_s^2 
    \left( \tilde{b} \ln\frac{M_s}{M'_c} + \Delta \right),
\label{eq:as-formula}
\end{equation}
at leading order in $\delta\alpha_s$.  An important point is that
$\delta\alpha_s$ is dominated by the first logarithmic term.
As we discussed before, threshold corrections from both $M_s$ and $M_c$ 
are under control and actually represented by $\Delta = O(1)$. Since 
$\ln(M_s/M'_c) \simeq 5/d$ for $C=5$, we find that $\delta\alpha_s$ is 
reliably estimated by knowing the $\beta$-function coefficients 
$\tilde{b}_i$, especially when $d=1$.

Let us now calculate $\tilde{b}_i$ in various type IV models. This can be 
done easily by using a diagrammatic technique. 
A remarkable thing is that $\tilde{b}_i$ do not depend on the detailed 
structure of the models.  They depend on only two things: the discrete 
symmetry used to define an orbifold and the higher dimensional multiplet 
containing the low-energy Higgs doublets. The basic idea is the following.
Consider an orbifold $M/K$, where $M$ is a manifold and $K$ is a discrete
group with $n_K$ elements. Then, we find a close relationship between 
the set of KK modes of $M/K$ and those of $M$: for each non-zero mode 
of $M/K$, there are $n_K$ corresponding modes of $M$ that are taken 
into each other by the elements of $K$. (For the example $M=S^1$ and 
$K=Z_2$: while $S^1$ has two states $e^{\pm i ny/R}$ at each non-zero 
energy, only one linear combination, $\cos (ny/R)$ or $\sin (ny/R)$, 
is available for a field on $S^1/Z_2$, since it must be 
either $+$ or $-$ under the $Z_2$ orbifold symmetry $y \rightarrow -y$.)
Hence, apart from the zero modes, the contribution to the gauge coupling 
running from some KK tower on $M/K$ is a fraction $1/n_K$ of the 
contribution from the corresponding KK tower on $M$.  However, we know 
that KK towers on $M$ produce only universal running, since $M$ has 
no fixed points at which local $G$ violation may occur (type I theories). 
This means that the non-universal running on $M/K$ is simply caused by 
a ``mismatch'' of the zero modes between the two towers: the zero 
mode contribution on $M/K$ is not necessarily $1/n_K$ of that on $M$.
This observation allows a very simple result for the values of 
$\tilde{b}_i$ on $M/K$: if some gauge component ${\cal U}$ of 
a ($4+d$)-dimensional supermultiplet has the zero mode on $M/K$, 
it contributes to $\tilde{b}_i$ by
\begin{equation}
  \tilde{b}_i = b_i^{{\cal U}_0} - \frac{1}{n_K} b_i^{\cal U},
\label{eq:general}
\end{equation}
where $b_i^{{\cal U}_0}$ are the 4D $\beta$-function coefficients from 
the zero mode, while $b_i^{\cal U}$ are those from the excited KK level 
of ${\cal U}$.\footnote{
In type II theories, the ``mismatch'' consists of a sum of KK towers 
characteristic of manifolds with dimensions less than $d$.
In general, we can easily read off full gauge coupling running equations, 
including the coefficients of non-universal power-law pieces, from 
this decomposition.}

To see how this works explicitly, let us consider 5D models on $S^1/Z_2$ 
in which gauge group $G$ is broken by orbifold boundary conditions. 
We first consider a gauge multiplet ${\cal V}^A = \{ V^A, \Sigma^A \}$, 
where $A$ is the gauge index and fields in the curly bracket represent 
4D $N=1$ superfields ($V$ and $\Sigma$ are vector and chiral 
superfields in the adjoint representation, respectively).
According to the transformation property under translation 
$y \rightarrow y+2\pi R$, the gauge multiplet ${\cal V}^A$ 
is divided into two classes: ${\cal V}^a(y+2\pi R) = {\cal V}^a(y)$ 
with KK masses $m_n = n/R$ and ${\cal V}^{\hat{a}}(y+2\pi R) = 
-{\cal V}^{\hat{a}}(y)$ with KK masses $m_n = (n+1/2)/R$, where $n$ 
takes integer values with $n \geq 0$.  Now, consider relating 
this KK tower to the corresponding KK tower on $S^1$, as discussed 
in the previous general analysis.  Suppose each KK level of 
${\cal V}^{\hat{a}}$ contributes to the running of $g_i$ with 4D 
$\beta$-function coefficients $b_i = b_i^X$. Then, as far as the runnings 
of $g_i$ are concerned, this KK tower, $(m_n, b_i) = ((n+1/2)/R), b_i^X)$ 
($n \geq 0$), is equivalent to the tower ${\cal T}^X: 
(m_n, b_i) = ((n+1/2)/R), b_i^X/2)$ ($-\infty < n < \infty$) that 
would be obtained on $S^1$.  The same rearrangement is possible for 
${\cal V}^a$, but in this case there is an extra subtlety coming 
from the presence of zero modes: the 4D $\beta$-function coefficients 
of the zero modes, $b_i^{A_0}$, are in general different from 
half of those of the excited modes, $b_i^A$.  
Thus the equivalent pattern for ${\cal V}^a$ consists of the tower 
${\cal T}^A: (m_n, b_i) = (n/R, b_i^A/2)$ ($-\infty < n < \infty$) and 
an additional ``effective zero mode'' with $b_i = b_i^{A_0} - b_i^A/2$.
This explicitly shows that the KK tower on $S^1/Z_2$ is equivalent 
to the tower $\{ {\cal T}^X, {\cal T}^A \}$ on $S^1$, apart from a 
slight mismatch represented by the ``effective zero mode''.
Since the pattern of the tower, $\{ {\cal T}^X, {\cal T}^A \}$, is 
completely the same as that obtained on $S^1$ (type I theories), 
they contribute only to $G$-symmetric power-law piece with 
$b^{(1)} = (b_i^X+b_i^A)/2$, which means that the $G$-violating piece 
entirely comes from the ``effective zero mode'' and is given by 
$\tilde{b}_i = b_i^{A_0} - b_i^A/2$.  This provides an explicit example 
of the general result, Eq.~(\ref{eq:general}), with $n_K = 2$.

The result can be simplified further by observing that there are simple 
relations between $b_i^{A_0}$ and $b_i^A$, which depend only on 
zero mode representations; if the zero modes come from $V$ and $\Sigma$, 
they are given by $b_i^A = (2/3)b_i^{A_0}$ and $b_i^A = -2 b_i^{A_0}$, 
respectively.  Therefore, we finally obtain the simple result for the 
contribution to $\tilde{b}_i$ from the gauge multiplet: 
$\tilde{b}_i = (2/3) b_i^{A_0}$ and $2 b_i^{A_0}$ in the case of 
$V$ and $\Sigma$ zero modes, respectively.
The contribution from a hypermultiplet, ${\cal H} = \{\Phi, \Phi^c\}$, 
can be worked out similarly ($\Phi$ and $\Phi^c$ represent 
4D $N=1$ chiral superfields).  The result is given by 
$\tilde{b}_i = b_i^{H_0} - b_i^H/2$, where $b_i^{H_0}$ and $b_i^H$ are 
the 4D $\beta$-function coefficients of the zero modes and the 
excited modes, respectively.  However, since $b_i^H = 2b_i^{H_0}$ 
in the hypermultiplet case, we find that it does not contribute to 
$\tilde{b}_i$, $\tilde{b}_i = 0$, if the extra dimension is $S^1/Z_2$.

The extension to the case with $d > 1$ is straightforward.
In the case of $d = 2$, the KK pattern is plotted in the two-dimensional 
momentum space $(p_5, p_6)$.  The only extra complication, compared with 
the case of $d=1$, is that the original orbifold KK towers do not 
necessarily fill half of this momentum plane; if the orbifold is 
$T^2/Z_m$, they fill only $1/m$ of the plane.  By repeating 
a similar analysis to the $d=1$ case, we obtain 
$\tilde{b}_i = (1-2/3m) b_i^{A_0}, (1+2/m) b_i^{A_0}$ and 
$(1-2/m) b_i^{H_0}$ in the case of $V$, $\Sigma$ and $\Phi$ zero modes, 
respectively.  Since the difference comes only from the fraction of 
momentum plane filled by KK towers, the $d=1$ case is reproduced by 
setting $m=2$ in these expressions.

With the above knowledge of $\tilde{b}_i$'s, we can calculate 
$\tilde{b}$ from Eq.~(\ref{eq:tilde-b}) in any type IV theories.  
In our framework the massless sector consists of the MSSM states, 
so that we only have to consider the KK excitations 
for these states.  First, the matter fields do not contribute to 
$\tilde{b}$ even if they live in the bulk, since they fill complete 
$SU(5)$ multiplets and so are the excited states 
($\tilde{b}_1 = \tilde{b}_2 = \tilde{b}_3$).  The gauge fields come from 
$V \subset {\cal V}$, so their contribution is given by 
$(\tilde{b}_1, \tilde{b}_2, \tilde{b}_3) = (0, -6+4/m, -9+6/m)$, using 
$(b_1^{A_0}, b_2^{A_0}, b_3^{A_0}) = (0, -6, -9)$.
For the Higgs doublets $H_u$ and $H_d$, there are three possibilities: 
they can originate from $\Phi \subset {\cal H}$ or $\Sigma \subset {\cal V}$ 
\cite{Hall:2001zb}, or can be brane fields ${\cal B}$ localized on 
a non-universal fixed point \cite{Hebecker:2001wq}. 
In each case, their contribution is given by 
$(\tilde{b}_1, \tilde{b}_2, \tilde{b}_3) = (3/5-6/5m, 1-2/m, 0)$,
$(\tilde{b}_1, \tilde{b}_2, \tilde{b}_3) = (3/5+6/5m, 1+2/m, 0)$, and 
$(\tilde{b}_1, \tilde{b}_2, \tilde{b}_3) = (3/5, 1, 0)$, 
respectively.  Adding all together, we finally obtain the values of 
$\tilde{b}$ for general type IV theories.  The result is summarized in 
Table.~\ref{tab:tilde-b}.
\begin{table}
\begin{center}
\begin{tabular}{|c|c|c|c|} \hline
  & $H_u, H_d \subset {\cal H}$ & $H_u, H_d \subset {\cal V}$ 
    & $H_u, H_d = {\cal B}$ 
\\ \hline
  $S^1/Z_2$ & $6/7$   & $-12/7$  & $-3/7$  \\
  $T^2/Z_m$ & $12/7m$ & $-24/7m$ & $-6/7m$
\\ \hline
\end{tabular}
\end{center}
\caption{The values of $\tilde{b}$ in general type IV theories, where 
 $H_u$ and $H_d$ represent two Higgs doublets in the MSSM.  The values for 
 the $S^1/Z_2$ case correspond to setting $m=2$ in the $T^2/Z_m$ case.}
\label{tab:tilde-b} 
\end{table}
Since the derivation is general, $T^2$ could be replaced by any 
two-dimensional compact manifold, and $Z_m$ by any discrete group 
with $m \geq 2$ elements.

We now have a list for $\delta\alpha_s$ in all type IV theories, 
which must be compared with the value, $\alpha_s^{\rm exp} - 
\alpha_s^{\rm SGUT,0} \simeq -0.013 \pm 0.004$. Here we used 
$\alpha_s^{\rm SGUT,0}$ to represent the value obtained including full 
two-loop effects, $\alpha_s^{\rm SGUT,0} \simeq 0.130 \pm 0.004$.  Using 
$M_s/M'_c \simeq \pi (300/C)^{1/d}$, we calculate $\delta\alpha_s$ 
for $C=5$ from the first term of Eq.~(\ref{eq:as-formula}).
For $d=1$, $\delta\alpha_s \simeq (-0.010, +0.020, +0.005)$ and for 
$d=2$ $\delta\alpha_s \simeq (-0.012/m, +0.024/m, +0.006/m)$ for the 
cases of ($H_{u,d} \subset {\cal H}$, $\subset {\cal V}$, $= {\cal B}$).
Does the logarithmic running between $M_c$ and $M_s$ resolve the 
discrepancy between $\alpha_s^{\rm SGUT,0}$ and $\alpha_s^{\rm exp}$? 
In the cases that the Higgs doublets originate from a ($4+d$)-dimensional 
gauge multiplet or are brane fields the answer is clearly no, since the 
negative sign of $\tilde{b}$ leads to an even larger discrepancy. The Higgs
must be bulk hypermultiplet fields, and furthermore, a precise agreement 
with data is only possible for the single case of $d=1$.\footnote{
In principle, we could add more bulk hypermultiplets if they have brane 
mass terms giving the zero modes a mass of order $M_c$.  For $m=2$, 
which includes all theories with $d=1$, they do not contribute to 
$\tilde{b}_i$ and thus do not affect the values of $\delta\alpha_s$ or 
$M_c$ obtained here. For $d=2, m>2$ (and for some of type III theories), 
these additional multiplets could contribute to $\tilde{b}_i$. While we 
cannot exclude these more complicated theories, they are ad hoc, and 
reminiscent of fixing up non-supersymmetric theories by populating the 
desert with additional split multiplets. The same is true for adding 
extra multiplets on branes with local $G$ violation.}

The case of $d=2$ and low $m$ cannot be excluded, although it is
certainly not preferred. For example, an $SO(10)$ theory on $T^2/Z_2$ 
($d=2,m=2,C=8$) \cite{Hall:2001xr} leads to a central prediction of 
$\alpha_s \simeq 0.124$ at leading logarithm. Measurements of the 
superpartner masses will determine whether such theories are excluded 
or not. They require very characteristic supersymmetry breaking parameters 
such as squarks and gluinos above 1 TeV and/or highly non-universal 
gaugino masses.  It is significant that the theories which are excluded 
or disfavored by gauge coupling unification generically pose problems for 
model building: models with the Higgs originating from gauge multiplets 
have difficulties in obtaining sizable Yukawa couplings (or must have 6D 
$N=2$ supersymmetry, but then they are Type III theories); if the Higgs 
doublets are on the brane then they are typically not subject to charge 
quantization; and if $d=2$ it is generically difficult to satisfy 
stringent constraints from anomaly cancellation in the 6D bulk.

\begin{figure}
\begin{center} 
\begin{picture}(350,250)(-10,-100)
  \LongArrow(0,-100)(0,150) \Text(-5,150)[r]{coupling}
  \Line(-4,125)(4,125) \Text(-10,128)[r]{$\approx 4\pi$}
  \Line(-4,45)(4,45) \Text(-10,48)[r]{$\approx 1$}
  \LongArrow(-20,0)(320,0) \Text(320,-5)[t]{energy}
  \DashLine(40,0)(40,140){4} \Text(40,-4)[t]{$M_Z$}
  \DashLine(180,-90)(180,140){4} 
  \Text(180,-4)[t]{$M_c$} \Text(177,-16)[t]{$\approx 10^{15}~{\rm GeV}$}
  \DashLine(250,-90)(250,140){4}
  \Text(250,-4)[t]{$M_s$} \Text(247,-16)[t]{$\approx 10^{17}~{\rm GeV}$}
  \Line(250,125)(180,39) \Line(180,39)(40,48)
  \Text(35,52)[r]{$\hat{g}_3$}
  \Line(250,125)(180,35) \Line(180,35)(40,33)
  \Text(35,37)[r]{$\hat{g}_2$}
  \Line(250,125)(180,31) \Line(180,31)(40,23)
  \Text(35,25)[r]{$\hat{g}_1$}
  \DashLine(250,125)(200,-25){2}
  \DashLine(200,-25)(180,-65){2}
  \DashLine(180,-65)(160,-95){2}
  \Text(155,-95)[r]{$\hat{g}_G$}
  \Text(110,115)[b]{4D $N=1$}
  \Text(110,100)[b]{$SU(3) \times SU(2) \times U(1)$}
  \Text(110,85)[b]{(MSSM)}
  \Text(215,130)[b]{5D $N=1$}
  \Text(215,115)[b]{$SU(5)$}
  \Text(315,85)[b]{strongly coupled}
  \Text(315,70)[b]{10D string (?) theory}
  \LongArrow(250,-85)(200,-85)
  \Text(225,-80)[b]{gravitational} \Text(225,-86)[t]{bulk (?)}
\end{picture}
\caption{The picture presenting our framework.  Here, we have plotted 
 $\hat{g}_i$ defined by $\hat{g}_i \equiv g_i(\mu/M_c)^{1/2}$ ($g_i$)
 for $\mu > M_c$ ($\mu < M_c$), which represent actual strengths 
 of the three gauge interactions.  The strength of the gravitational 
 interaction, $\hat{g}_G \equiv (\mu/M_{*D})^{(D-2)/2}$, is also plotted, 
 where $D$ is the spacetime dimension in which gravity propagates and 
 $M_{*D}$ the reduced Planck scale in $D$ dimensions.}
\label{fig:frame}
\end{center}
\end{figure}
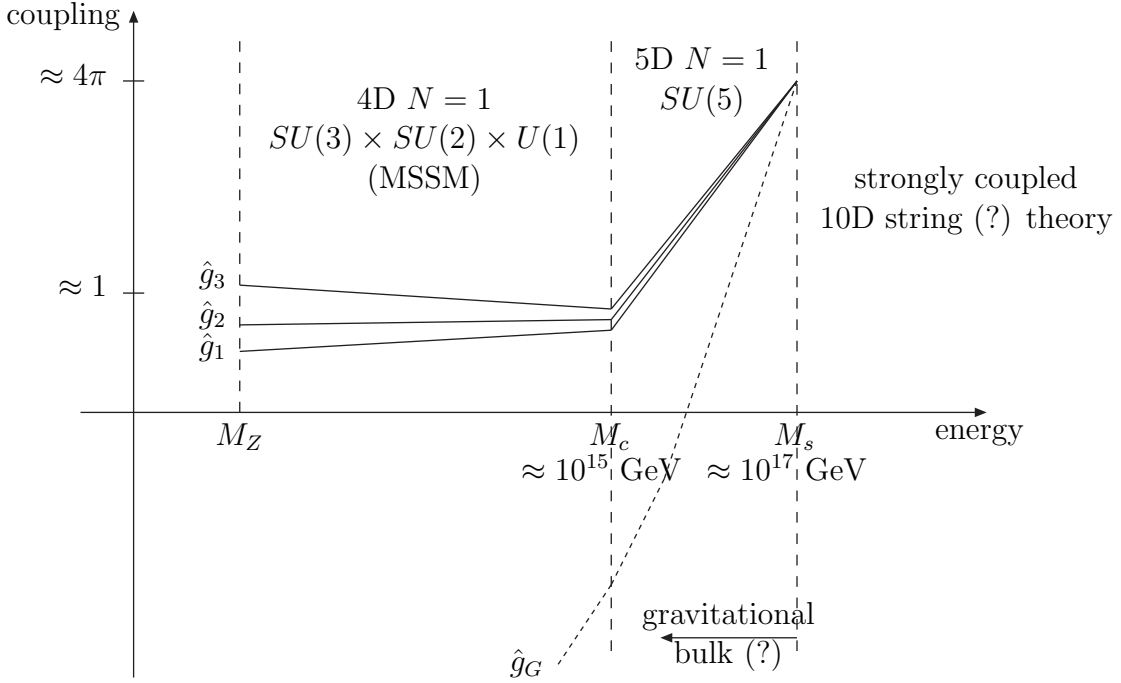

We conclude that our framework strongly favors $d=1$, and therefore
the gauge group $G = SU(5)$, since it is the largest group 
that can be broken to the standard model gauge group 
by compactifying on $S^1/Z_2$.  (Note that larger gauge groups are 
also disfavored from the fact that they have larger values of $C$ and 
thus lead to smaller values of $\ln(M_s/M'_c)$ and $|\delta\alpha_s|$.)
The bulk matter content is also fixed to be two Higgs hypermultiplets, 
up to a possibility of putting matter in the bulk, since otherwise there 
remain unwanted massless fields at low energies.
The compactification scale $M_c$ is calculated using Eq.~(\ref{eq:running})
as $M_c = \pi M_u (M'_c/M_s)^{5/7} \approx 10^{15}~{\rm GeV}$.
Therefore, we finally arrive at the following picture:  
there is a large energy interval ranging from $10^{15}$ to 
$10^{17}~{\rm GeV}$ in which the physics is described by a higher 
dimensional grand unified field theory, and it must actually be a 5D 
$SU(5)$ theory with the two Higgs hypermultiplets propagating in the bulk.
The overall physical picture is summarized in Fig.~\ref{fig:frame}.

Here we make one brief comment.  In section \ref{sec:framework} we 
defined our framework to have $d$ extra dimensions of comparable radii. 
It is important to note that there is no need for the radii to be 
comparable, rather this was a simplifying assumption. For example, 
to obtain a tree-level gauge coupling unification at $M_s$, we require 
only that the volume of the bulk is large, and much larger than the 
volume of non-universal branes.  The addition of extra radii will 
result in extra parameters entering the prediction for the QCD gauge 
coupling, so that it is possible to get a continuum of results, 
interpolating between the discrete possibilities that follow for 
a single radius.  For example, 6D $SO(10)$ theories compactified on 
an asymmetric space, $R_5 \neq R_6$, can interpolate $\alpha_s$ 
predictions from that of $d=2$ to that of $d=1$, and therefore can 
agree well with data.  Nevertheless, it is still true that the best fit 
value of $\alpha_s$ is obtained for $R_5 \gg R_6$, in which case the 
effective theory in the energy region between $R_5^{-1}$ and $R_6^{-1}$ 
is 5D $SU(5)$ ($\times U(1)$), and the physics picture is 
almost that of Fig.~\ref{fig:frame} with $M_c \rightarrow R_5^{-1}$ and 
$M_s \approx R_6^{-1}$.  Obviously, in this case, what we previously 
said as a fixed point can actually be a fixed line in 6D with 
a sufficiently small volume $V' \approx 1$.

Having obtained a detailed picture of the structure of the theory 
around the unified scale, we here discuss uncertainties for the 
present analysis.  First, there is an uncertainty for the ratio 
$M_s/M_c$ due to a universal power-law running of $g_i$ above $M_c$.
However, its effect on $\delta\alpha_s$ is small; for instance, even 
a factor 2 uncertainty in $M_s/M_c$ gives only about 15\% uncertainty 
in $\delta\alpha_s$.  Second, there are higher loop field theory 
contributions just below $M_s$, which are no longer loop suppressed 
since the theory is becoming strongly coupled.  However, this strongly 
coupled physics is occurring only over a very small energy interval, and 
the theory is weakly coupled in most of the energy region between 
$M'_c$ and $M_s$ where the logarithmic contribution comes from.  This is 
because in the 5D picture the couplings in the theory have negative 
mass dimensions, and in the 4D picture the number of KK states circulating 
in the loop decreases with decreasing energies so that loop expansion 
parameters in the theory become small by powers of $(\mu/M_s)$.
Therefore, we expect that these contributions are no larger than those 
from the physics above $M_s$ encoded in the operators localized 
on the non-universal fixed points \cite{Nomura:2001tn}.  We estimate 
the resulting uncertainty to be about $\ln(3)/\ln(60\pi) \simeq 20\%$, 
assuming that the strong coupling physics makes the one-loop value of 
$\tilde{b}$ unreliable above $M_s/3$.  There are also uncertainties 
in $\Delta$ which represents the threshold corrections from $M_s$ 
and $M_c$.  However, since we have identified the matter content 
of the theory, the latter contribution becomes a calculable quantity; 
if we adopt dimensional regularization, for example, it is 
given by $\Delta_{M_c} \simeq 0.84$ \cite{Contino:2001si}.
Thus, here we take $\Delta = 0.84 \pm 1$, where the error represents 
the threshold correction from $M_s$ and a renormalization scheme 
dependence.  Adding up all together, we finally obtain the value 
$\delta\alpha_s = -0.012 \pm 0.003$, which is translated into 
the prediction of our framework given in Eq.~(\ref{eq:alphas.kk}).
It is important to notice that the difference, $\delta\alpha_s$, 
from the conventional one-scale unification comes dominantly from the 
logarithmic evolution between $M'_c$ and $M_s$.  Unlike the threshold 
correction $\Delta_{\rm SGUT}$ in the usual framework, there is no free 
parameter which can be chosen to adjust $\delta\alpha_s$; the masses 
and couplings for the particles around the unification scale are all 
determined by the orbifold compactification.  We have achieved 
a significant improvement over the conventional supersymmetric grand 
unification framework.

\section{The $SU(5)$ Model in Five Dimensions}
\label{sec:su5}

We have shown that our framework, a higher dimensional unified theory
breaking on an orbifold, correctly predicts the QCD coupling only in the 
unique situation that the orbifold is $S^1/Z_2$ and the unified gauge 
symmetry is $SU(5)$. In this section we discuss features of this 5D 
$SU(5)$ theory, showing that other aspects are also determined.
We explicitly present a completely realistic theory, which illustrates 
how various problems in 4D supersymmetric grand unified theories are 
elegantly solved in the KK grand unification framework.

The orbifold boundary conditions are unique. The orbifold reflection 
of ${\cal Z}: y \rightarrow -y$ preserves $SU(5)$, having parities 
for gauge and hypermultiplets of ${\cal V} = \{ V(+), \Sigma(-) \}$ 
and ${\cal H} = \{\Phi(+), \Phi^c(-) \}$. Under translation 
${\cal T}: y \rightarrow y + 2 \pi R$, the $SU(5)$ is broken by the 
action of $P = \mbox{diag}(+,+,+,-,-)$ on a 5-plet.\footnote{
This is equivalent to the boundary conditions of 
Ref.~\cite{Kawamura:2001ev} described in terms of ${\cal Z}$ and 
${\cal Z}' = {\cal Z}{\cal T}$ as $S^1/(Z_2 \times Z_2')$.}
In addition, bulk hypermultiplets can have extra factors 
$\eta_{\Phi} = \pm 1$ under the translation.  These are the most 
general boundary conditions preserving 4D $N=1$ supersymmetry.
Specifically, the boundary conditions for the 
gauge and hypermultiplets are written as
\begin{equation}
  \pmatrix{V^{(\pm)} \cr \Sigma^{(\pm)}}(x^\mu,y) 
  = \pmatrix{V^{(\pm)} \cr -\Sigma^{(\pm)}}(x^\mu,-y) 
  = \pm \pmatrix{V^{(\pm)} \cr \Sigma^{(\pm)}}(x^\mu,y+2\pi R), \\
\label{eq:bc-g}
\end{equation}
\begin{equation}
  \pmatrix{\Phi^{(\pm)} \cr \Phi^{(\pm)c}}(x^\mu,y) 
  = \pmatrix{\Phi^{(\pm)} \cr -\Phi^{(\pm)c}}(x^\mu,-y) 
  = \pm\, \eta_{\Phi} 
    \pmatrix{\Phi^{(\pm)} \cr \Phi^{(\pm)c}}(x^\mu,y+2\pi R), 
\label{eq:bc-h}
\end{equation}
where we have labelled the standard model gauge multiplets and 
the broken $SU(5)$ gauge multiplets as 
$(V^{(+)},\Sigma^{(+)})=(V_{321},\Sigma_{321})$ and 
$(V^{(-)},\Sigma^{(-)})=(V_X,\Sigma_X)$, respectively; 
$\Phi^{(+)}$ ($\Phi^{(-)}$) represents the components of $\Phi$ that 
are even (odd) under the action of $P$.

We have shown that the two Higgs doublets of the MSSM must arise from 
bulk hypermultiplets, rather than as 
fields localized at the $y = \pi R$ brane, which respects only
the standard model gauge symmetry.  Thus we introduce two Higgs 
hypermultiplets $\{ H,H^c \}$ and $\{ \bar{H},\bar{H}^c \}$ in the bulk, 
which transform as ${\bf 5}$ and $\bar{\bf 5}$ under $SU(5)$.
These Higgs multiplets must have $\eta_H = \eta_{\bar{H}} = -1$ to have 
massless Higgs doublets.  (In the present notation, $H^{(+)}$ and 
$H^{(-)}$ ($\bar{H}^{(+)}$ and $\bar{H}^{(-)}$) represent triplet and 
doublet components, $H_T$ and $H_D$ ($\bar{H}_T$ and $\bar{H}_D$), 
of $H$ ($\bar{H}$), respectively.)  Then, from 
Eqs.~(\ref{eq:bc-g}, \ref{eq:bc-h}), we obtain the following 
KK towers for the gauge and Higgs fields: the standard model 
gauge vectors $V_{321}$ with masses $n/R$, joined at $n\neq 0$ 
levels by the $N=2$ partners $\Sigma_{321}$; the broken 
$SU(5)$ vectors $V_X$, joined by their $N=2$ partners $\Sigma_X$, 
of mass $(n + 1/2)/R$; two Higgs doublets $H_D$ and $\bar{H}_D$ with 
masses $n/R$, joined at $n\neq 0$ levels by the $N=2$ partners 
$H^c_D$ and $\bar{H}^c_D$; and two Higgs triplets $H_T$ and $\bar{H}_T$ 
with masses $(n+1/2)/R$ joined by their $N=2$ partners $H^c_T$ and 
$\bar{H}^c_T$.  These KK towers are summarized in Table~\ref{table:Z-T}. 
The Higgs KK towers do not have zero modes for the color triplet 
states \cite{Kawamura:2001ev}, and the KK excitations do not lead to 
proton decay from dimension five operators because their mass term takes 
the form $H H^c + \bar{H} \bar{H}^c$ rather than 
$H \bar{H}$ \cite{Hall:2001pg}.  It is precisely these vector and Higgs 
towers which lead to the successful prediction for the QCD coupling.
\begin{table}
\begin{center}
\begin{tabular}{|c|c|c|c|}
\hline
 $({\cal Z},{\cal T})$  &  gauge and Higgs fields  & 
    bulk matter fields & 4D masses \\ \hline
 $(+,+)$  & $V_{321}$,      $H_D$,   $\bar{H}_D$   & 
    $T_{U,E}$, $T'_Q$,     $F_D$, $F'_L$     & $n/R$       \\ 
 $(+,-)$  & $V_{X}$,        $H_T$,   $\bar{H}_T$   & 
    $T_Q$, $T'_{U,E}$,     $F_L$, $F'_D$     & $(n+1/2)/R$ \\ 
 $(-,+)$  & $\Sigma_{321}$, $H^c_D$, $\bar{H}^c_D$ & 
    $T^c_{U,E}$, $T'^c_Q$, $F^c_D$, $F'^c_L$ & $(n+1)/R$ \\ 
 $(-,-)$  & $\Sigma_{X}$,   $H^c_T$, $\bar{H}^c_T$ & 
    $T^c_Q$, $T'^c_{U,E}$, $F^c_L$, $F'^c_D$ & $(n+1/2)/R$   \\ 
\hline
\end{tabular}
\end{center}
\caption{The transformation properties for the bulk fields 
under the orbifold reflection and translation.  Here, we have used 
the 4D $N=1$ superfield language.  The fields written in the 
$({\cal Z},{\cal T})$ column, $\varphi$, obey the boundary condition 
$\varphi(y) = {\cal Z}\varphi(-y) = {\cal T}\varphi(y+2\pi R)$.
The masses for the corresponding KK towers are also given 
($n=0,1,\cdots$).}
\label{table:Z-T}
\end{table}

To preserve the $SU(5)$ understanding of matter quantum numbers 
the quarks and leptons should either be in the bulk or reside 
at the $SU(5)$ preserving brane at $y=0$ \cite{Hall:2001pg}. 
Yukawa interactions are forbidden by 5D supersymmetry from appearing
in the bulk Lagrangian, and hence must be brane localized.  If quarks 
and leptons are on the brane, they fill out 4D chiral multiplets 
which are ${\bf 10}$ or $\bar{\bf 5}$ representations of 
$SU(5)$: $T$ and $F$.  The Yukawa interactions are located 
on the $y=0$ brane as
\begin{equation}
  S = \int d^4x \; dy \; \delta(y)
    \biggl[ \int d^2\theta \Bigl( y_T T T H + y_F T F \bar{H} \Bigr) 
    + {\rm h.c.} \biggr].
\label{eq:yukawa-1}
\end{equation}
Since the full $SU(5)$ symmetry is operative at $y=0$, we have $SU(5)$ 
mass relations for the quarks and leptons localized on the brane.
The resulting 4D Yukawa couplings are suppressed by a factor of 
$1/(M_s R)^{1/2}$ due to the Higgs wavefunctions being spread out 
over the bulk.  On the other hand, if quarks and leptons are in the bulk, 
they arise from hypermultiplets: $\{ T,T^c \} + \{ T',T'^c \}$ and
$\{ F,F^c \} + \{ F',F'^c \}$ with $\eta_T=\eta_F=1$ and 
$\eta_{T'}=\eta_{F'}=-1$.  We find from Eq.~(\ref{eq:bc-h}) that 
a generation $q,u,d,l,e$ arise from the zero modes of bulk 
fields as $T(u,e), T'(q), F(d)$ and $F'(l)$.  (Note that 
$T^{(+)} = T_{U,E}$, $T^{(-)} = T_{Q}$, $F^{(+)} = F_{D}$, 
$F^{(-)} = F_{L}$, and similarly for $T'$ and $F'$, where $T_{Q,U,E}$ 
($F_{D,L}$) are the components of $T$ ($F$) decomposed into irreducible 
representations of the standard model gauge group. The tower structure 
for these fields is given in Table~\ref{table:Z-T}.)
Since $q$ and $u,e$ ($d$ and $l$) come from different hypermultiplets, 
the broken gauge boson exchange does not lead to proton decay.
The Yukawa couplings are introduced on the $y=0$ brane as
\begin{eqnarray}
  S &=& \int d^4x \; dy \; \delta(y)
    \biggl[ \int d^2\theta \Bigl( y_T^1 T T H + y_T^2 T T' H + y_T^3 T' T' H 
\nonumber\\
    && + y_F^1 T F \bar{H} + y_F^2 T F' \bar{H} + y_F^3 T' F \bar{H} 
    + y_F^4 T' F' \bar{H} \Bigr) + {\rm h.c.} \biggr].
\label{eq:yukawa-2}
\end{eqnarray}
Although these are $SU(5)$ symmetric interactions, the quarks and 
leptons do not respect $SU(5)$ mass relations because the 
down-type quark and charged lepton masses come from $y_F^3$ and $y_F^2$ 
couplings, respectively, and are not related by the $SU(5)$ 
symmetry.\footnote{
We could also introduce Yukawa couplings that do not respect the 
$SU(5)$ symmetry, on the $y=\pi R$ brane.}
Moreover, since the matter wavefunctions are also spread out in the 
extra dimension, the resulting 4D Yukawa couplings receive a stronger 
suppression, a factor of $1/(M_s R)^{3/2}$, than in the case of 
brane matter.  Thus we find a clearly successful correlation between 
the mass of the fermion and whether it has $SU(5)$ mass relations 
--- heavier fermions display $SU(5)$ mass relations while 
lighter ones do not.  Of course, if we have both bulk and brane matter, 
we can also write down the Yukawa couplings that mix them, 
on the $y=0$ brane.

We now discuss an important issue of what brane localized operators 
can be introduced in our theory.  The 5D restricted gauge symmetry 
alone allows many unwanted operators on the branes.  For instance, 
the operators $[H \bar{H}]_{\theta^2}$ and $[F H]_{\theta^2}$ give 
a large mass, of order the unified scale, for the Higgs doublets 
destroying the solution to the doublet-triplet splitting problem, 
$[T F F]_{\theta^2}$ causes disastrous dimension four proton decay, and 
$[Q Q Q L]_{\theta^2}$ induces too rapid dimension five proton decay. 
In addition, if matter is located in the bulk, we could also 
have $SU(5)$ non-invariant operators on the $y=\pi R$ brane, such as 
$[T_Q T_Q \bar{H}^c_T]_{\theta^2}$ and $[T_Q F_L H^c_T]_{\theta^2}$, 
which reintroduce the problem of dimension five proton decay caused by 
colored Higgsino exchange.  Remarkably, however, the structure of the 
theory allows a mechanism that simultaneously suppresses all these 
unwanted operators \cite{Hall:2001pg}.  Since the bulk Lagrangian has 
higher dimensional supersymmetry, it possesses an $SU(2)_R$ symmetry.
It also has an $SU(2)_H$ flavor symmetry rotating the two Higgs 
hypermultiplets in the bulk.  After the orbifolding, these two $SU(2)$ 
symmetries are broken to two $U(1)$ symmetries, one from 
$SU(2)_R$ and one from $SU(2)_H$.  A particularly interesting 
symmetry is the diagonal subgroup of these $U(1)$ symmetries, which 
we call $U(1)_R$ symmetry since it is an $R$ symmetry rotating the 
Grassmann coordinate of the low energy 4D $N=1$ supersymmetry.
We can extend this bulk $U(1)_R$ symmetry to the full theory 
by assigning appropriate charges to the brane localized 
quark and lepton superfields, and use it to constrain possible forms 
of brane localized operators.  The resulting $U(1)_R$ charges are 
given in Table~\ref{table:U1R}, where $T$ and $F$ (and $N$) represent 
both brane and bulk matter.  Imposing this $U(1)_R$ symmetry on the 
theory, we can forbid unwanted operators while keeping the Yukawa 
couplings.  The dimension four and five proton decays are prohibited, 
and the $R$-parity violating operators are absent since $U(1)_R$ 
contains the usual $R$ parity as a discrete subgroup.  After
supersymmetry breaking, this $U(1)_R$ symmetry is broken (presumably 
to its $R$-parity subgroup), generating gaugino masses and the 
supersymmetric mass term for the two Higgs doublets ($\mu$ term) 
of the order of the weak scale.  Since the breaking scale is small, 
however, it will not reintroduce the problem of proton decay.  It is 
interesting to note that the spacetime symmetries of the theory allows 
a bulk mass term of the form $[H \bar{H} - H^c \bar{H^c}]_{\theta^2}$, 
coupling the two Higgs hypermultiplets. This would remove the Higgs 
doublets from the low energy theory and reintroduce dimension five 
proton decay from colored Higgsino exchange. The $U(1)_R$ symmetry also 
forbids this bulk mass term, providing a complete solution to the 
doublet-triplet and proton decay problems.
\begin{table}
\begin{center}
\begin{tabular}{|c|cc|cccc|cccccc|}  \hline 
  & V & $\Sigma$ & $H$ & $H^c$ & $\bar{H}$ & $\bar{H}^c$ 
  & $T$ & $T^c$ & $F$ & $F^c$ & $N$ & $N^c$ \\ \hline
  $U(1)_R$ & 0 & 0 & 0 & 2 & 0 & 2 & 1 & 1 & 1 & 1 & 1 & 1 \\ \hline
\end{tabular}
\end{center}
\caption{$U(1)_R$ charges for 4D vector and chiral superfields.}
\label{table:U1R}
\end{table}

We here comment on neutrino masses.  Small neutrino masses are 
generated through the see-saw mechanism \cite{Seesaw}, if we introduce 
right-handed neutrino superfields.  They could be either brane fields, 
$N$, or bulk fields, $\{ N,N^c \}$ with $\eta_N=1$.  The Yukawa 
couplings, $[F N H]_{\theta^2}$, and Majorana masses, $[N N]_{\theta^2}$, 
are written on the brane.  The $U(1)_R$ charges for these fields 
are given in Table~\ref{table:U1R}.

Now we ask how we can determine the location of matter fields.
Since our framework gives $M_c \approx 10^{15}$ GeV, the $X$ gauge 
bosons are considerably lighter, of mass about $10^{15}$ GeV, than 
in the case of 4D supersymmetric grand unification.  This makes 
dimension six proton decay a non-trivial issue in our theory.
We find that the quarks and leptons of the first generation coming 
from a ${\bf 10}$ representation must be bulk fields, since 
otherwise the $X$ gauge boson exchange would induce proton decay 
at too rapid a rate.\footnote{
The authors of Ref.~\cite{Contino:2001si} did not consider the 
possibility of bulk matter, and hence concluded that unification in 
5D did not improve the 4D unification prediction for the QCD coupling.}
We will say that $T_1$ is in the bulk, although we really mean the
combination $\{ T_1,T^c_1 \} + \{ T'_1,T'^c_1 \}$.
On the other hand, the top quark must arise from a brane field $T_3$.
If the top quark were a bulk mode, it would have a mass 
suppressed by a factor of $1/(M_s R)^{3/2}$, which gives too light 
a top quark even in the case that the Yukawa coupling is strong. 
With $T_3$ on the brane, strong coupling leads to a top Yukawa coupling 
of the low energy theory of $4 \pi /(M_s R)^{1/2} \approx 1$, 
giving a top quark mass of the observed size.  Thus we are able to 
derive the location of both the first and third generation ${\bf 10}$'s, 
and we find that at least some aspects of flavor physics are 
associated with the geometry of the orbifold, and with strong coupling. 
Arguments can be made for the location of the rest of the quarks and 
leptons, although these are not strict requirements.  For example, 
the rest of the third generation, $F_3$, is best placed on the brane, 
giving the successful $SU(5)$ mass prediction for $m_b/m_\tau$. 
On the other hand, some (or all) of lighter generations are located 
in the bulk so that it does not exhibit unwanted $SU(5)$ mass relations.

For the lighter two generations we mention two interesting possibilities.
The large $\nu_\mu \nu_\tau$ mixing, observed in atmospheric neutrino
fluxes, suggests that $F_2 \supset \nu_\mu$ is also on the brane
so that it has a large mixing with $F_3 \supset \nu_\tau$. With these
assignments the location of the rest of the second generation is
fixed: $T_2$ must be in the bulk, otherwise all the second and third 
generation fermions would be on the brane, leading to an incorrect 
$SU(5)$ mass prediction between the strange quark and the muon.  A bulk 
location for $T_2$ is in any case desired, since it leads to small CKM 
mixing between second and third generations, $V_{cb} \approx \epsilon$, 
and to a mass hierarchy $m_\mu/m_\tau, m_s/m_b \approx \epsilon$ and 
$m_c/m_t \approx \epsilon^2$. It is interesting to note that,
in the case that all the Yukawa couplings of the heaviest two
generations are strongly coupled, $\epsilon \approx (M_c/M_s)^{1/2}
\approx 0.1$, so that all the above relations are good at the factor of
3 level. This requires the large $m_t/m_b$ ratio to result from a
large ratio of electroweak vacuum expectation values, $\tan\beta$.
The only remaining question is the location of $F_1$, which is not
constrained by proton decay. One possibility is that all three $F_i$'s
are on the brane. One might expect this to give large angle solar
neutrino oscillations.  However, such a location implies that the first
two generations are not distinguished by their spatial location. Thus
the hierarchies of the masses and mixings of the first two generations
must come from elsewhere. Another possibility is for $F_1$ to be in
the bulk. In this case the hierarchies $m_d/m_s, m_e/m_\mu \approx
\epsilon$, but the smallness of the Cabibbo angle, and, more
importantly, of $m_u/m_c$ are not explained.  It appears that many, 
but not all, aspects of flavor can be understood from this single
extra dimension \cite{Nomura:2001tn,Hall:2001rz}.

Another possibility is to have two generations on the brane and one in 
the bulk. The bulk generation must be interpreted as $T_1,F_2$, rather 
than $T_1,F_1$, which would lead to an incorrect relation for $m_s/m_\mu$. 
The extra dimension cannot explain all aspects of flavor --- some 
additional ingredient is needed. In the present case a very simple 
approximate flavor symmetry is sufficient. The brane fields, with 
flavor charges in parentheses, are $T_3(0), F_3(1), T_2(1), F_1(1)$, 
while the bulk fields are $T_1(1), F_2(0)$. The size of entries 
in the Yukawa matrices are determined by a combination of factors 
of the volume of the bulk, $\epsilon$, and the size of the flavor 
symmetry breaking parameter, $\delta$.  In the case that 
$\epsilon \approx \delta$, it gives the following Yukawa matrices:
\begin{equation}
  {\cal L}_4 \approx 
  \pmatrix{
     T_1 & T_2 & T_3 \cr
  }
  \pmatrix{
     \epsilon^4 & \epsilon^3             & \epsilon^2           \cr
     \epsilon^3 & \underline{\epsilon}^2 & \underline{\epsilon} \cr
     \epsilon^2 & \underline{\epsilon}   & \underline{1}        \cr
  }
  \pmatrix{
     T_1 \cr T_2 \cr T_3 \cr
  } H
\nonumber\\
  + \epsilon
  \pmatrix{
     T_1 & T_2 & T_3 \cr
  }
  \pmatrix{
     \epsilon^2           & \epsilon^2 & \epsilon^2           \cr
     \underline{\epsilon} & \epsilon   & \underline{\epsilon} \cr
     \underline{1}        & 1          & \underline{1}        \cr
  }
  \pmatrix{
     F_1 \cr F_2 \cr F_3 \cr
  } \bar{H},
\end{equation}
where only underlined entries must respect $SU(5)$ relations.
This well reproduces the qualitative pattern of the observed quark and 
lepton masses and mixings: $m_t:m_c:m_u \approx 1:\epsilon^2:\epsilon^4$,
$m_b:m_s:m_d \approx m_\tau:m_\mu:m_e \approx 1:\epsilon:\epsilon^2$, 
$(V_{us},V_{cb},V_{ub}) \approx (\epsilon,\epsilon,\epsilon^2)$, 
$m_b/m_\tau \simeq 1$, and $m_s/m_\mu \neq 1$.  The neutrino mixing angles 
are expected to be bi-maximal, giving large angle solutions to the solar 
neutrino problem.  Unlike our earlier example, the present matter 
configuration has a local cancellation of gauge anomalies 
and does not need a Chern-Simons term. The two examples also have very 
different power-law running of the 4D gauge coupling above $M_c$. With 
$T_2$ in the bulk this couplings blows up at about $40 M_c$; this reduces 
$\ln(M_s/M'_c)$ below our central value, but gives a prediction for the 
QCD coupling within our quoted uncertainty. With $T_2$ on the brane, the 
one-loop coefficient of the power-law running vanishes. Of course, the 5D 
coupling is still strong at $M_s$.

The location of $T_2$ is very important for gauge boson mediated
proton decay, which only occurs via brane localized $T_i$.  In our 
first flavor model, $T_2$ is in the bulk, so that proton stability 
is expected ($\tau_p \approx 10^{39}-10^{41}~{\rm years}$). 
In the second model, $T_2$ is on the brane, leading to proton decay 
in the Cabibbo suppressed channels $\mu^+ K^0$ and $K^+ \bar{\nu}_\tau$, 
at an interesting rate for future experiments 
($\tau_p \approx 10^{33}-10^{35}~{\rm years}$).

\section{Relation to String Theory}
\label{sec:string}

Superstring theories are formulated in 10D and therefore require 
compactification on a 6D space.  Most work on compactification has 
concentrated on a 6D space which is close to symmetrical, with the 
six radii all comparable in size. Also much attention has been paid 
to perturbative heterotic string theory. The message of this paper is
that string theory should be strongly coupled, and compactified on 
a highly asymmetric space, with one radius, $R$, much larger
than the others, $\tilde{R}$. The mass scale of the strongly coupled 
string theory is near $10^{17}$ GeV.\footnote{
Another possibility is that the string theory is just perturbative, 
but close to being strongly coupled. In the case of the heterotic 
$E_8 \times E_8'$ theory this is not possible, as the string scale 
is $(\sqrt{\alpha}/2) M_{\rm Pl} \approx 10^{18}$ GeV, an order of 
magnitude larger than $M_s$.  Our theory may be realized 
in brane world scenarios.}

The value of the unified gauge coupling requires the volume of the 6D
compact space to be $\simeq 60$, using fundamental units of $M_s$. 
Much of this volume will arise from the large dimension, so that 
$\ln(M_s/M_c)$ is large enough to correct the usual supersymmetric 
prediction of $\alpha_s$.  However, the other radii need not be exactly 
$1/M_s$; the present uncertainty from the superpartner thresholds allows 
for $\tilde{R}$ to be somewhat larger than $1/M_s$.  Measurements of 
superpartner masses would place tighter restrictions on this.

At distances larger than $\tilde{R}$, the 5D effective theory has 
$SU(5)$ gauge interactions propagating in spacetime $M^4 \times S^1/Z_2$. 
Bulk modes include hypermultiplets for two ${\bf 5}$ of Higgs
fields, and for (${\bf 10,10'}$) which has the lightest generation
as zero modes. The remaining matter fields may be bulk or brane
modes, except for the top quark, which must be contained in a brane
${\bf 10}$. Unlike attempts to get 4D grand unified theories from string
theory, there should not be adjoint or other fields for breaking the
unified gauge symmetry. Rather there must be an $SU(5)$ breaking twist
in the translation boundary condition for $S^1$.  It will be interesting 
to pursue 10D string models which reduce to the above 5D $SU(5)$ theory 
below $\tilde{R}^{-1}$.\footnote{
An interesting intermediate step for this construction might be 
a 6D $N=2$ $SU(6)$ model along the line of Ref.~\cite{Hall:2001zb}, 
compactified on a $T^2/(Z_2 \times Z'_2)$ orbifold with 
$R_5 \gg R_6 \approx M_s^{-1}$.  Imposing the boundary condition, 
${\cal Z}_5 = \mbox{diag}(+,+,+,+,+,+),{\cal Z}_6 = \mbox{diag}(+,+,+,+,+,-),
{\cal T}_5 = \mbox{diag}(+,+,+,-,-,-),{\cal T}_6 = \mbox{diag}(+,+,+,+,+,+)$
acting on ${\bf 6}$ of $SU(6)$, the structure of the theory precisely 
reduces to that of the $SU(5)$ theory discussed here (with an additional 
$U(1)$) below $R_6^{-1}$.  The $SU(6)$ gauge multiplet in 6D reproduces 
the 5D $SU(5)$ gauge multiplet (plus $U(1)$) with two Higgs hypermultiplets 
in a ${\bf 5}$ representation.  The matter fields are located on 
the $(x_5,x_6)=(0,0)$ fixed point or the $x_6=0$ fixed line.
The Yukawa couplings are introduced on the $(x_5,x_6)=(0,0)$ brane.}

There may be several ways to realize our framework in string 
theory. One possibility is the strongly coupled $E_8 \times E_8'$ 
heterotic string theory, which can be viewed as a 11D supergravity theory
having a large ``gravity-only'' dimension \cite{Horava:1996qa}. 
The resulting 11D theory can be written, using standard notation, as
\begin{equation}
  S = \int d^4x d^6y dz \left\{ \frac{1}{2\kappa^2} \sqrt{g} {\cal R}
    - \frac{1}{8\pi(4\pi\kappa^2)^{2/3}} \sqrt{g} \left( 
    \delta(z){\rm tr}F^2 + \delta(z-\pi\rho){\rm tr}F'^2 \right) \right\}.
\end{equation}
The eleventh dimension $z$ has a radius $\rho$ of size
\begin{equation}
  \frac{1}{\rho} = (8\pi^2 \alpha^{-1})^{3/2}
    \left( \frac{1}{V^{1/2} M_{\rm Pl}^2} \right).
\end{equation}
where $\alpha \simeq 1/24$ is the unified gauge coupling.  
For a symmetrical 6D space, with six comparable radii \cite{Horava:1996qa}, 
$V \approx M_u^{-6}$ and $1/\rho \simeq 4 \times 10^{15}$ GeV. 
In our framework, the large asymmetry in the 6D compact space implies 
$V \approx M_s^{-5} M_c'^{-1} \approx M_u^{-6} (M'_c/M_s)^{5/7}$, 
so that $1/\rho$ is increased to about $9 \times 10^{15}$ GeV. 
In summary: if our framework is described by strongly coupled heterotic 
string theory, the fundamental string scale is close to $10^{17}$ GeV, 
five dimensions of the compact space are close to this scale, but 
two have much larger radii, characterized by the mass scales 
$10^{16}$ GeV and $10^{15}$ GeV, respectively. They are both described 
by $S^1/Z_2$, but the former is a ``gravity-only'' dimension, while the 
latter allows propagation of $SU(5)$ gauge interactions.

\section{Discussion and Conclusions}
\label{sec:concl}

\begin{figure}
\begin{center} 
\begin{picture}(275,300)(-40,0)
  \LongArrow(0,0)(0,300) \Text(-5,300)[r]{$\alpha_s$}
  \Line(-3,15)(3,15)   \Text(-8,16)[r]{$0.060$}
  \Line(-3,45)(3,45)
  \Line(-3,75)(3,75)   \Text(-8,76)[r]{$0.080$}
  \Line(-3,105)(3,105)
  \Line(-3,135)(3,135) \Text(-8,136)[r]{$0.100$}
  \Line(-3,165)(3,165)
  \Line(-3,195)(3,195) \Text(-8,196)[r]{$0.120$}
  \Line(-3,225)(3,225)
  \Line(-3,255)(3,255) \Text(-8,256)[r]{$0.140$}
  \Line(-3,285)(3,285)
  \Text(210,186)[l]{$\alpha_s^{\rm exp}$}
  \DashLine(0,180)(200,180){4}
  \DashLine(0,192)(200,192){4}
  \Text(50,285)[b]{$\alpha_s^{\rm GUT}$}
  \DashLine(50,33)(50,99){2}
  \DashLine(45,33)(55,33){2} \DashLine(45,99)(55,99){2} 
  \DashLine(50,60)(50,72){1} \Vertex(50,66){3}
  \DashLine(45,60)(55,60){1} \DashLine(45,72)(55,72){1} 
  \Text(100,285)[b]{$\alpha_s^{\rm SGUT}$}
  \DashLine(100,192)(100,258){2}
  \DashLine(95,192)(105,192){2} \DashLine(95,258)(105,258){2}
  \DashLine(100,216)(100,234){1}
  \DashLine(95,216)(105,216){1} \DashLine(95,234)(105,234){1}
  \Line(100,213)(100,237) \Vertex(100,225){3}
  \Line(95,213)(105,213) \Line(95,237)(105,237)
  \Text(150,285)[b]{$\alpha_s^{\rm KK}$}
  \DashLine(150,180)(150,198){1}
  \DashLine(145,180)(155,180){1} \DashLine(145,198)(155,198){1}
  \Line(150,177)(150,201) \Vertex(150,189){3}
  \Line(145,177)(155,177) \Line(145,201)(155,201)
\end{picture}
\caption{The predictions for $\alpha_s$ in the three frameworks: 
 non-supersymmetric grand unification $\alpha_s^{\rm GUT}$, 
 supersymmetric grand unification $\alpha_s^{\rm SGUT}$, and 
 Kaluza-Klein grand unification $\alpha_s^{\rm KK}$. 
 Solid error bars represent the threshold corrections from the 
 superpartner spectrum. Dotted error bars for $\alpha_s^{\rm GUT}$ and 
 $\alpha_s^{\rm SGUT}$ represent threshold corrections from
 the unified scale corresponding to a heavy ${\bf 5} + \bar{\bf 5}$
 representation with unit logarithmic mass splitting between doublets
 and triplets. The dashed error bars represent possible dependence 
 on models from physics at the unification scale: particle content, 
 higher dimensional operators, coupling constants, etc, and have been
 arbitrarily normalized to bring $\alpha_s^{\rm SGUT}$ in agreement
 with experiment. The dotted error bar for $\alpha_s^{\rm KK}$ is the
 theoretical uncertainty (other than from superpartner masses) for 
 our theory, as estimated in the text.}
\label{fig:alpha-s}
\end{center}
\end{figure}
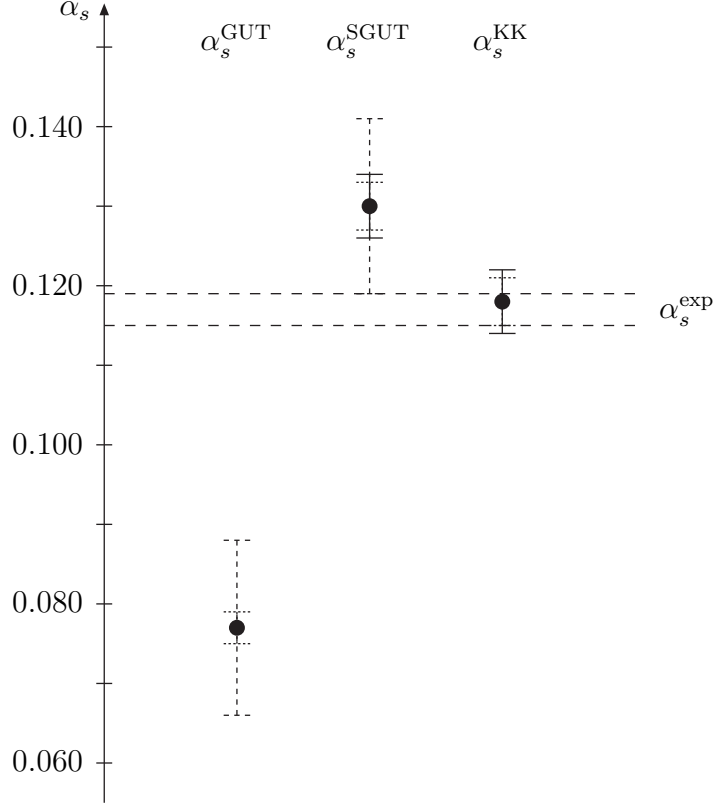

The QCD, weak and electromagnetic forces play very different roles in
nature, and, at first sight, it seems very unlikely that they
are all manifestations of a single unified interaction. Nevertheless,
the structure of the standard model does allow an elegant picture of
unification, although the resulting prediction for the QCD coupling is
50\% from the observed value, as shown in Fig.~\ref{fig:alpha-s}. 
This picture of unification introduces several problems into the 
structure of the theory. First, the unification occurs at the enormous 
energy scale of $10^{15}$ GeV, introducing a large hierarchy with the 
weak scale. Actually, there is already a hierarchy between the weak scale 
and the Planck mass, the scale at which gravity becomes strong, and it is, 
perhaps, disappointing that the unification scale is fully four orders 
of magnitude lower than the Planck mass.  Other problems include 
excessive proton decay induced by the unified gauge bosons, breaking the 
unified gauge symmetry, and understanding why the Higgs doublets and 
their color triplet partners have a hierarchical mass splitting.

Weak scale supersymmetry has successfully addressed some of these
problems, so that the MSSM up to a very high energy scale has become
the standard paradigm for new physics. The superpartners of the gauge 
and Higgs bosons lead to a marked improvement in the prediction 
for the QCD coupling, as shown in Fig.~\ref{fig:alpha-s}. 
At the same time the unification scale is raised 
to $2 \times 10^{16}$ GeV, removing the problem of gauge boson mediated 
proton decay, and diminishing the distance to the Planck scale. 
The superpartners also lead to a radiative stability of the hierarchy 
of mass scales.  Despite these successes, an energy desert of 13 orders 
of magnitude is a startling conclusion, and should not be drawn lightly. 
There are scenarios without a desert: for example, large extra 
dimensions can lead to gravity getting strong at the TeV scale
\cite{Arkani-Hamed:1998rs}, and string theory can occur at the TeV scale 
\cite{Lykken:1996fj}. There are many weak arguments against a low
fundamental mass scale; for example, from proton decay, neutrino
masses and inflation. The only argument which has real strength, since
it is based on the numerical prediction of a measured quantity, is 
that of gauge coupling unification.  On this score the low scale 
theories do poorly: typically they cannot yield a simple picture of 
coupling unification. Such a picture does exist in the case
of power-law unification in higher dimensions \cite{Dienes:1998vh}, but
the accuracy of the prediction is greatly weakened through high sensitivity 
to unknown ultraviolet physics. In the case of accelerated unification 
\cite{Arkani-Hamed:2001vr}, an accurate prediction persists,
but at the cost of multiple replications of the standard model gauge group.
Nevertheless, it must be admitted that high scale gauge coupling unification
is not perfect. The central value for the prediction of the QCD
coupling is about 10\% off, requiring large threshold corrections from
the unified scale. For example, if superheavy ${\bf 5} + \bar{\bf 5}$
chiral multiplets of $SU(5)$ are added, with a unit logarithmic mass
splitting between doublet and triplet components, the threshold correction 
is small, $\Delta_{\rm SGUT} \simeq 0.003$, as shown by the dotted 
uncertainty drawn in Fig.~\ref{fig:alpha-s} for $\alpha_s^{\rm SGUT}$. 
The correction required by data, shown by the dashed error bar in 
Fig.~\ref{fig:alpha-s}, is much larger. Furthermore, in supersymmetric 
theories the questions of breaking the unified gauge group and the mass 
splitting between Higgs doublets and triplets still remain, and the 
further problem of proton decay from dimension five operators is 
introduced. However, these are objections against 4D grand unified 
theories rather than high scale gauge coupling unification.

The discrepancy between the experimental values of the gauge couplings 
and the prediction from supersymmetric unification is usually ascribed 
to threshold corrections from the unified scale which depend on unknown 
parameters or moduli of the unified theory. In this paper we have 
taken an alternative viewpoint. We have discovered a new framework
that offers the possibility of a reliably calculated, precision agreement 
with experimental data. The discrepancy of the standard supersymmetric 
prediction is accounted for by a moderately large logarithmic effect 
in a higher dimensional unified theory, with orbifold breaking of the 
gauge symmetry. The size of this logarithm is determined by the 
strong coupling requirement.  Remarkably, this framework does correctly 
predict the central experimental value for the QCD coupling, as shown in 
Fig.~\ref{fig:alpha-s}. Furthermore, threshold corrections from the scale 
of the unified gauge boson masses are unambiguous and have been included. 
The remaining uncertainty from unknown physics at even higher energies 
can be reliably estimated to be small, as shown by the dotted error bar 
of Fig.~\ref{fig:alpha-s} for $\alpha_s^{\rm KK}$.

Since the framework allows for many possible models, each with differing 
coefficients of the logarithm, it is fair to question whether we have 
really predicted the data, or whether we have used the data to select 
a model. We view the situation as somewhat analogous to the case of 
supersymmetric unification. Within that framework there are many possible 
models, for example ones with $2n$ Higgs doublets, each giving a different 
prediction for the QCD coupling. Nevertheless, the addition of weak scale 
supersymmetry is viewed as highly significant because the simplest 
possible model is precisely the one that works best. All the more 
complicated theories are very much further from the data. We have found 
a similar situation to hold in the case of adding extra dimensions at the
unified scale.  The various models lead to a discrete set of predictions, 
yet it is only the simplest model, with one extra dimension and $SU(5)$ 
gauge group, that is able to precisely account for the data; the majority 
of models find a correction which is either too small or of the wrong 
sign. As precision electroweak measurements strengthened the case for the 
MSSM, future measurements of superpartner masses will further test minimal 
KK unification. In supersymmetric unification the crucial new running 
is induced by one set of superpartners for the minimal set of gauge and 
Higgs bosons. Adding extra dimensions, we find that precision unification 
follows from the running induced by one set of KK modes for the 
minimal gauge and Higgs bosons. In our view, this observation 
strengthens the case for a high fundamental scale.

In this paper we have studied an alternative to 4D grand unification 
or string theory for physics just beyond the supersymmetric desert 
which leads to gauge coupling unification: Kaluza-Klein grand 
unification \cite{Hall:2001pg}.  We have introduced a new framework 
which predicts the leading radiative corrections to gauge coupling 
unification from the high scale.  There is an essentially unique theory 
which provides a precise and successful prediction for the QCD coupling. 
Not only does this improve on the prediction from conventional 
supersymmetric unification, but it also solves the three outstanding 
problems of 4D grand unified theories. The color triplet partners of 
the doublet Higgs bosons are projected out of the zero mode sector 
by the orbifold boundary condition, the underlying $R$ symmetry 
of the theory automatically removes all proton decay from dimension 
four and five operators, and light fermions are guaranteed not to have 
$SU(5)$ mass relations.

Furthermore, the addition of extra dimensions leads to new avenues of 
exploration for flavor. We find that the top quark is necessarily 
a brane mode, while part of the first generation is necessarily 
in the bulk. Some aspects of flavor must be associated with the geometry 
of the extra dimension. It is interesting that at the fundamental 
scale the top Yukawa interaction, and perhaps other flavor couplings, 
are strongly coupled.  Finally, it is intriguing to note that the size 
of the neutrino mass suggested by atmospheric neutrino oscillations is 
related to the compactification scale, $M_c$, by $v^2/M_c$, where $v$ 
is the electroweak vacuum expectation value.

\section*{Acknowledgements}

Y.N. thanks the Miller Institute for Basic Research in Science 
for financial support.  This work was supported in part by the Director, 
Office of Science, Office of High Energy and Nuclear Physics, of the U.S. 
Department of Energy under Contract DE-AC03-76SF00098, and in part 
by the National Science Foundation under grant PHY-00-98840.

\end{document}